\documentclass{article} % For LaTeX2e
\usepackage{iclr2021_conference,times}

\usepackage{amsmath,amsfonts,bm}

\def\eqref#1{equation~\ref{#1}}
\def\1{\bm{1}}

\DeclareMathAlphabet{\mathsfit}{\encodingdefault}{\sfdefault}{m}{sl}
\SetMathAlphabet{\mathsfit}{bold}{\encodingdefault}{\sfdefault}{bx}{n}

\usepackage{hyperref}
\usepackage{url}
\usepackage{amsmath,amssymb,amsfonts}
\usepackage{mathtools}
\usepackage{algorithmic}
\usepackage{graphicx}
\usepackage{textcomp}
\usepackage{xcolor}
\usepackage{xspace}
\usepackage{multicol}
\usepackage{multirow}
\usepackage{url}
\usepackage{listings}
\usepackage[utf8]{inputenc}
\usepackage{bm}
\usepackage{subcaption}
\usepackage{gensymb}
\usepackage{makecell}
\usepackage{etoolbox}
\usepackage{soul}
\usepackage{tikz}
\usepackage[ruled,vlined]{algorithm2e}
\usepackage{pgf-umlsd}
\usepackage{balance}
\usepackage{float}
\usepackage{wrapfig}

\newcommand{\ignore}[1]{}
\usepackage{ifthen}
\newboolean{showcomments}
\setboolean{showcomments}{true} %false true
\ifthenelse{\boolean{showcomments}}{
\newcommand{\mynote}[3]{\noindent{\color{#3}\textbf{#1\xspace} #2}}}
{ \newcommand{\mynote}[3]{}
}
\definecolor{mypurple}{rgb}{0.6,0.4,0.8}
\definecolor{mygreen}{rgb}{0.31,0.61,0.12}

\definecolor{myblue}{RGB}{0, 114, 178} % Hex color #0072B2
\definecolor{myorange}{RGB}{230, 159, 0} % Hex color #E69F00
\definecolor{mygreen}{RGB}{0, 158, 115} % Hex color #009E73
\definecolor{mygrey}{RGB}{153, 153, 153} % Hex color #999999

\DeclareRobustCommand\blueline
{\tikz[baseline=0.0ex]\draw[myblue, line width=1pt] (0,0.1) -- (0.35,0.1);}
\DeclareRobustCommand\orangeline
{\tikz[baseline=0.0ex]\draw[myorange, line width=1pt] (0,0.1) -- (0.35,0.1);}
\DeclareRobustCommand\greenline
{\tikz[baseline=0.0ex]\draw[mygreen, line width=1pt] (0,0.1) -- (0.35,0.1);}

\DeclareRobustCommand\blackdashedline
{\tikz[baseline=0.0ex]\draw[black, dashed, line width=1pt] (0,0.1) -- (0.35,0.1);}

\newcommand{\ie}{\textit{i.e.}\xspace}
\newcommand{\eg}{\textit{e.g.}\xspace}
\newcommand{\wrt}{\textit{w.r.t.}\xspace}

\title{Layer-wise Characterization of Latent Information Leakage in Federated Learning}

\author{Fan Mo, Anastasia Borovykh, Mohammad Malekzadeh, Hamed Haddadi \& Soteris Demetriou\\
Imperial College London\\
\texttt{\{f.mo18,a.borovykh,m.malekzadeh,h.haddadi,s.demetriou\}@imperial.ac.uk}
}
\iclrfinalcopy

\begin{document}

\maketitle

\begin{abstract}
Training deep neural networks via federated learning allows clients to share, instead of the original data, only the model trained on their data. Prior work has demonstrated that in practice a client's private information, unrelated to the main learning task, can be discovered from the model's gradients, which compromises the promised privacy protection. However, there is still no formal approach for quantifying the leakage of private information via the shared updated model or gradients. In this work, we analyze property inference attacks and define two metrics based on (i) an adaptation of the \emph{empirical $\mathcal{V}$-information}, and (ii) a \emph{sensitivity analysis} using Jacobian matrices allowing us to measure changes in the gradients with respect to latent information. We show the applicability of our proposed metrics in localizing private latent information in a layer-wise manner and in two settings where (i) we have or (ii) we do not have knowledge of the attackers' capabilities. We evaluate the proposed metrics for quantifying information leakage on three real-world datasets using three benchmark models.
\end{abstract}
 
\section{Introduction}
Federated learning~(FL), allows {\em clients} to jointly train a model, \eg a deep neural network~(DNN), on their local data, and iteratively share their updates with a {\em server} that aggregates the received updates~\citep{mcmahan2017communication}. There is a surge of interest in FL~\citep{kairouz2019advances, bonawitz2017practical, hard2018federated}, as it resolves the need for collecting private data in a centralized location which in many cases is not possible due to computational costs, privacy risks, and even legal considerations~\citep{kairouz2019advances, bonawitz2019towards}. However, it is found that even sharing the gradients of private data \wrt a DNN's parameters is not a privacy panacea as these gradients can contain private information~\citep{melis2019exploiting, zhu2019deep, nasr2019comprehensive}. 

The {\em property inference attacks}~(PIA)~\citep{melis2019exploiting, wang2019beyond} in FL aim to infer some private information about a targeted client, \eg~a client's property like gender or race. Although there are some works~\citep{neyshabur2017exploring, lee2020finite, arora2019exact, achille2019information} on the memorization and generalization of DNNs \wrt~the main task that DNNs are trained on, these works are not able to explain the memorization of latent information independent of the main task. There also are some works on (layer-wise) understanding of the learned representation by a DNN~\citep{zeiler2014visualizing, mahendran2015understanding, shwartz2017opening, saxe2019information}, but their proposed methods only provide insights on how information evolves during \emph{forward propagation}, while the analysis of information captured in the \emph{backward propagation} is still missing.

{\bf Contributions.} We focus on the open problem of how PIA leverages latent information presented in the computed gradients, and show that the Shannon mutual information~\citep{shannon1948mathematical} cannot properly quantify the risk of PIA. We propose adopting a more generalized notion of information, \emph{$\mathcal{V}$-information}~\citep{xu2020theory}, also known as `usable' information. This can measure the layer-wise latent information privacy risk and explain which layers are the most vulnerable to PIA. Furthermore, as $\mathcal{V}$-information requires knowledge of the specific family of attack models, we further present a metric based on a \emph{sensitivity analysis} of the gradients \wrt~latent information, which is more suitable when modeling a general adversary. 

\section{Privacy analysis: metrics for quantifying property privacy}

% \section{The setting} 
% \paragraph{Property privacy.} 
{\bf Threat Model.} We aim at characterizing \emph{property privacy}~\citep{melis2019exploiting, hitaj2017deep, zhu2019deep}, which is one type of latent information privacy, as opposed to \emph{input privacy}~\citep{bonawitz2017practical, gu2018yerbabuena} that refers to the privacy of training data. Our focus is on the properties~(\eg gender or race) that can be inferred from input data~(\eg face images). As the property of a client in FL can even be inferred from the average gradients computed on the entire client's data, it is much more challenging to protect such latent information, compared to preventing pixel-wise reconstruction of the client's input data.
% As the property of data is high-level latent information which aggregated gradients can easily carry, the property privacy is more difficult to preserve than input privacy. 
% \paragraph{Adversary.}
We assume the PIA adversary $\mathcal{A}$~(\eg the server or a malicious client) aims to disclose the property of a targeted client (\ie victim) by observing gradients or updated models broadcast by the server~\citep{melis2019exploiting}.
%We consider an adversary present on a FL participating node (\ie~one client) who aims to disclose private information of other clients by observing updated gradients or global models broadcast from the server.
% Based on the updated gradients/models, $\mathcal{A}$ can infer 
We assume the client's private properties are irrelevant to the main task of FL. To conduct the attack, we assume $\mathcal{A}$ can observe multiple model updates released by the victim and $\mathcal{A}$ has access to some (public) auxiliary data.
% first needs to observe multiple snapshots of the global model from the server and collects other information including auxiliary data about the property. The auxiliary data can be its own data or public data so they are easy to collect.
Thus, $\mathcal{A}$ can train an \emph{attack model}, \eg a binary classifier~\citep{melis2019exploiting}, using the auxiliary data and the collected model updates, to discover the victim's property.
% which can be obtained by subtracting two sequential global models, as having the target property or not in corresponding inputs (see \citep{melis2019exploiting} for more details on PIAs). 

\paragraph{Shannon Information.}
A well-known metric for quantifying the information flow of the training dataset in a DNN's forward propagation is the Shannon mutual information~(MI)~\citep{shwartz2017opening, goldfeld2019estimating}. Let $\mathrm{X}$, $\mathrm{Y}$, and $\hat{\mathrm{Y}}$ be the input, the ground-truth label, and the output of the DNN, respectively. Let $\mathrm{T}_l$ denote the intermediate representation in layer $l$ (we refer Appendix~\ref{sec:notations} for notations). The MI between $\mathrm{X}$ (or $\mathrm{Y}$) and $\mathrm{T}_l$ satisfies the Data Processing Inequality~(DPI)~\citep{tishby2000information},
\begin{equation*}
\begin{split}
& I(\mathrm{X};\mathrm{T}_1) \geq I(\mathrm{X};\mathrm{T}_2) \cdots \geq I(\mathrm{X};\mathrm{T}_L) \geq I(\mathrm{X};\hat{\mathrm{Y}}), \\
& I(\mathrm{Y};\mathrm{X}) \geq I(\mathrm{Y};\mathrm{T}_1) \geq I(\mathrm{Y}; \mathrm{T}_2) \cdots \geq I(\mathrm{Y};\mathrm{T}_L)\geq I(\mathrm{Y};\hat{\mathrm{Y}}).
\end{split}
\end{equation*}
These inequalities correspond to the intuition that information should \emph{not} increase during the layer-by-layer forward propagation in DNNs. However, DPI fails to explain the backward propagation because the Markov chain construction on inputs, model parameters, and computed \emph{gradients} is much more complex than that in the forward propagation~(see Figure~\ref{fig:mkc} in Appendices). One can reasonably expect a DNN to extract features that are useful for the target task such that `usable' information for the task $\mathrm{Y}$ memorized in DNN parameters/gradients, \ie~what is relevant for PIAs, would be \emph{increasing} throughout the layers.
%This intuition has been verified by analyzing the forward propagation: intermediate representations (\eg~$\mathrm{T}$) of the earlier layers contain more generic features that are useful for many tasks, but later layers become progressively more specific to the details relevant to the desired task~\cite{bengio2013representation, goldfeld2019estimating}. However, there is no successful MI analysis directly on gradients produced in the backward propagation.

\paragraph{$\mathcal{V}$-Information.} As claimed in~\cite{xu2020theory}, Shannon MI, due to its assumption of unbounded computational power, fails to explain that the `usable information' in the late layers of a DNN  may be higher than in early layers. Such a definition of usable information can be highly relevant to understanding PIA. We, therefore, argue that the notion $\mathcal{V}$-information from \cite{xu2020theory} is better suited for analyzing the latent information captured in the gradients of each DNN's layer. Here, we formulate PIA in the notion of $\mathcal{V}$-information.%\footnote{\scriptsize{For more technical details on the definitions we refer to the original paper~\citep{xu2020theory}.}}. \vincent{removed for space}

Let the adversary, due to computational constraints, only has access to the attack models $f_{\mathcal{A}}$ from a specific predictive family $\mathcal{V}_{\mathcal{A}}$ (\eg~a family of random forest algorithm). 
Let $\mathrm{p}\in \{0,1\}$ be the (binary) property and $\mathrm{G}_l \in \mathbb{R}^{N_l \times N_{l-1}}$ be the layer $l$'s gradients, taking values in the sample space, where $N_l \times N_{l-1}$ is the dimensionality of used gradients. Let $\mathrm{g}_l$ denote an instance of $\mathrm{G}_l$ and $f_{\mathcal{A}}[\mathrm{g}_l]\in[0,1]$ be a probability measure on the value of $\mathrm{p}$ computed via the side information $\mathrm{g}_l$. Thus $f_{\mathcal{A}}[\mathrm{g}_l](\mathrm{p}_i)\in[0,1]$ denotes the probability of $\mathrm{p}=i$, for $i\in\{0,1\}$, that is achieved via the softmax of the $f_{\mathcal{A}}$'s output. %We assume that the predictive family satisfies the optional ignorance condition from Definition~\ref{defpredfam} in Appendix~\ref{sec:appV}. Under certain assumptions a DNN and a random forest (RF) can satisfy this.
Let $\mathcal{D}$ be a training dataset including samples of $\{\mathrm{g}_l, \mathrm{p}\}$.
Let 
%$\mathrm{p}_i \in \{0,1\}$ denote whether or not the property is present for data sample~$i$ and let 
$\mathrm{g}_{l,i}$ be the layer~$l$'s gradients for data sample $i$. The {\em empirical $\mathcal{V}$-information} from $\mathrm{G}_l$ to $\mathrm{p}$~(\ie information about $\mathrm{p}$ available in $\mathrm{G}_l$) is defined as
\begin{equation}
\begin{split}
    \hat{I}_{\mathcal{V}_{\mathcal{A}}}(\mathrm{G}_l\rightarrow \mathrm{p}; \mathcal{D}) = \inf_{f_{\mathcal{A}} \in \mathcal{V}_{\mathcal{A}}} \frac{1}{|\mathcal{D}|} \sum_{\mathrm{p}_i \in \mathcal{D}} -\log f_{\mathcal{A}}[\emptyset](\mathrm{p}_i) - \inf_{f \in \mathcal{V}_{\mathcal{A}}} \frac{1}{|\mathcal{D}|} \sum_{\mathrm{g}_{l,i},\mathrm{p}_i \in \mathcal{D}} -\log f_{\mathcal{A}}[\mathrm{g}_{l,i}](\mathrm{p}_i).
\end{split}
\label{eq:v_info_adv}
\end{equation}
%
% Let $z$ be the output of the attack model.  
% The density $f_{\mathcal{A}}[\mathrm{g}_{l,i}](\mathrm{p}_i)$ evaluated at $\mathrm{p}_i$ is then given by a softmax of the model output $f[\mathrm{g}_l]$. 
In Section~\ref{sec:eval}, we show how to use the empirical $\mathcal{V}$-information to measure how well a property can be predicted when the adversary has access to the gradients of a layer as the side information. 
%When $f_{\mathcal{A}}[\mathrm{g}_l]$ produces a perfect prediction with a loss equal to 0 on $\mathcal{D}$, the second item in Equation~\ref{eq:v_info_adv} is equal to 0 and consequently $\mathcal{V}_{\mathcal{A}}$ achieves the highest $\mathcal{V}$-information between $\mathrm{G}$ and $\mathrm{P}$. In the worst case, $f_{\mathcal{A}}[\mathrm{g}_l]$ produces a random prediction regardless of whether we use or not use side information $\mathrm{g}_l$ so the smallest $\mathcal{V}$-information is equal to 0. 

\paragraph{Sensitivity.}
A downside of $\mathcal{V}$-information is that it relies on certain assumptions on the adversary's power; in particular the predictive family $\mathcal{V}_{\mathcal{A}}$. This makes $\mathcal{V}$-information practically similar to computing the adversary's area under curve~(AUC). Therefore we also propose to directly compute privacy risk information for a \emph{general} attacker, using solely the gradient information without making explicit assumptions about the attackers' model. We utilize the Jacobian matrix of the gradients as a straightforward \emph{sensitivity} metric, %\footnote{\scriptsize{
similar to input-output Jacobian in~\cite{novak2018sensitivity} and~\cite{sokolic2017robust}, for measuring the general privacy risk on gradients \wrt~high-level features (\eg~property). Intuitively, if certain gradients are non-sensitive to the input, or the property of data, the success of the attack can be expected to be lower. We compute the Jacobian of gradients \wrt~the main task ${\mathrm{y}}$ and we expect the Jacobian of gradients \wrt~a private property $\mathrm{p}$ to be similar to that on $\hat{\mathrm{y}}$, since both of them are high-level features.

%For a PIA as in \cite{melis2019exploiting}, the success of the attack lies in the ability of the classifier to distinguish between whether a certain gradient was computed over data that did or did not have a certain property. This is due to the fact that in this case different input data or properties will not result in significantly different layer gradients. Training a classifier that would be highly successful in distinguishing between gradients with or without a certain property will then be non-viable. 

% The Jacobian of gradients \wrt~one output $\hat{\mathrm{Y}}$ is calculated. We expect the evaluation \wrt~$\mathrm{p}$ to be similar to that on $\hat{\mathrm{Y}}$, since both of them are high-level features, and the PIA can be achieved by having either the property task (\ie~$\mathrm{p}$) or the main task (\ie~$\hat{\mathrm{Y}}$) at the end of a DNN for multi-learning~\cite{melis2019exploiting}. 

We can use the Frobenius norm ($F$-norm) for matrices and the $1$-norm or the $\infty$-norm for vectors in order to capture adversaries with different capabilities. Because Jacobians are compared across layers, the magnitude and size of the layer's parameters can also have an impact on the comparison results so we normalize the \emph{norm of Jacobian} with $\mathrm{G}_l$'s size. 
Thus, given $K$ data samples, we compute the `Jacobian $p$-norm' averaged over the data samples as the privacy risk,
\begin{equation}
\begin{split}
      \frac{1}{K\psi_l} \sum_{k=1}^K \left\lVert \frac{\mathbf{J}^{(\mathrm{G})}_l(\hat{\mathrm{Y}}_k)}{\text{range}(\mathrm{g}_l(\hat{\mathrm{Y}}_k))}\right\rVert_p
\end{split}
\label{eq:jaco_p}
\end{equation}
where $\mathbf{J}_l^{(\mathrm{G})}(\hat{\mathrm{Y}}) = \frac{\partial \mathrm{g}_l(\hat{\mathrm{Y}})}{\partial \hat{\mathrm{Y}}} = \frac{\partial}{\partial \hat{\mathrm{Y}}} \left( \frac{\partial \ell(\hat{\mathrm{Y}}, \bm{\theta}_l)}{\partial \bm{\theta}_l} \right)$ and $\mathrm{g}_l(.)$ represents the function that produces layer $l$'s gradients $\mathrm{G}_l \in \mathbb{R}^{N_l \times N_{l-1}}$ with true output $\hat{\mathrm{Y}}$, $\ell(.)$ is the loss function over $\hat{\mathrm{Y}}$ and $\bm{\theta}$ (parameters of the complete model), so $\mathrm{g}_l(.)$ can be regarded as the partial derivative of $\ell(.)$ \wrt~layer $l$'s parameters $\bm{\theta}_l$ (\ie~backward propagation). Function
$\text{range}(.)$ returns the range of values in one vector (\ie~$\text{max}(.)-\text{min}(.)$), and $p = F, 1,\text{or}\ \infty$. Parameter  $\psi_l$ is the normalization factor based on the size of the Jacobian matrix of layer $l$ under each $p$. Thus,
$\psi_l = \sqrt{N_l \times N_{l-1}}, \;N_l \times N_{l-1}$, and $\;1$ for $p = F, \; 1, \;$ and $\infty$, respectively.
% used to balance the dimension differences of layers. %We introduce $\psi_l$ because the considered latent private information leakage~\cite{melis2019exploiting} happens when the adversary concludes a binary inference (\ie~with/without the property); thus the final Jacobian norm should be reflected in one point (\ie~normalized by the layer size under each $p$).

\begin{figure}[t]
    \centering
    \includegraphics[width=0.325\textwidth]{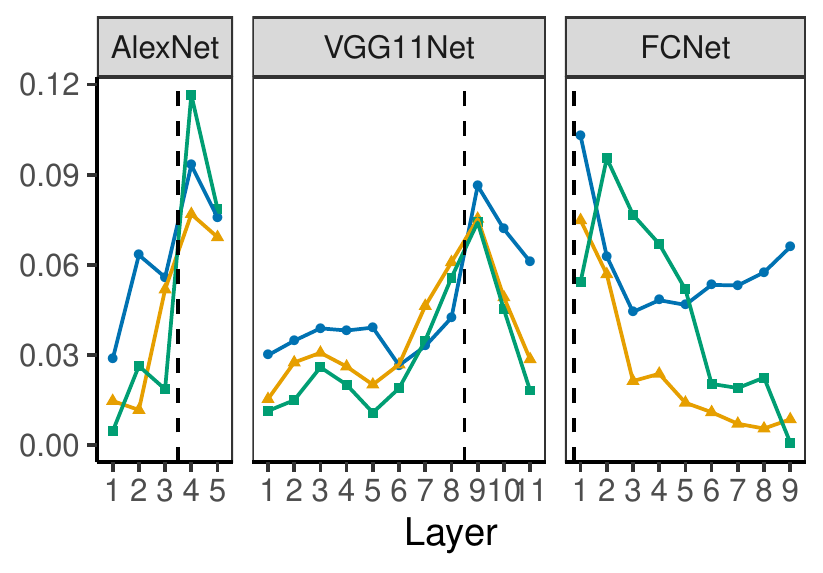}
    \includegraphics[width=0.325\textwidth]{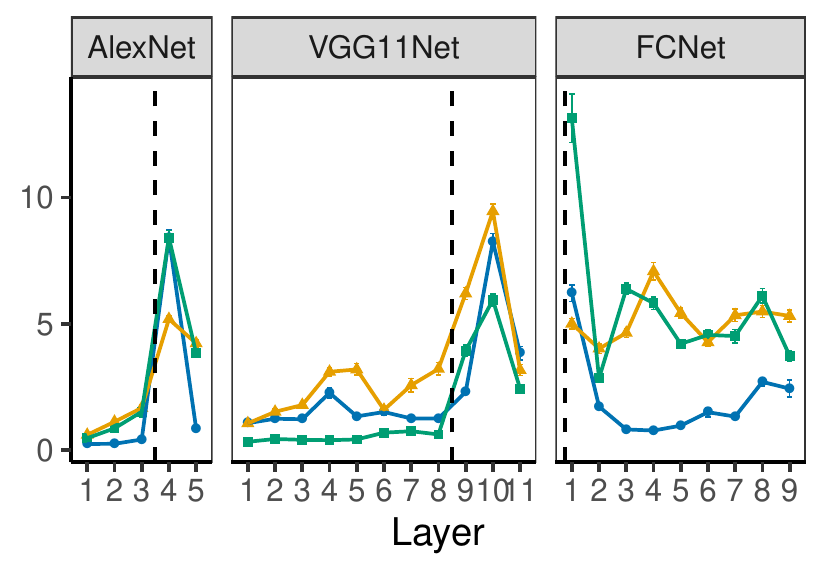}
    \includegraphics[width=0.325\textwidth]{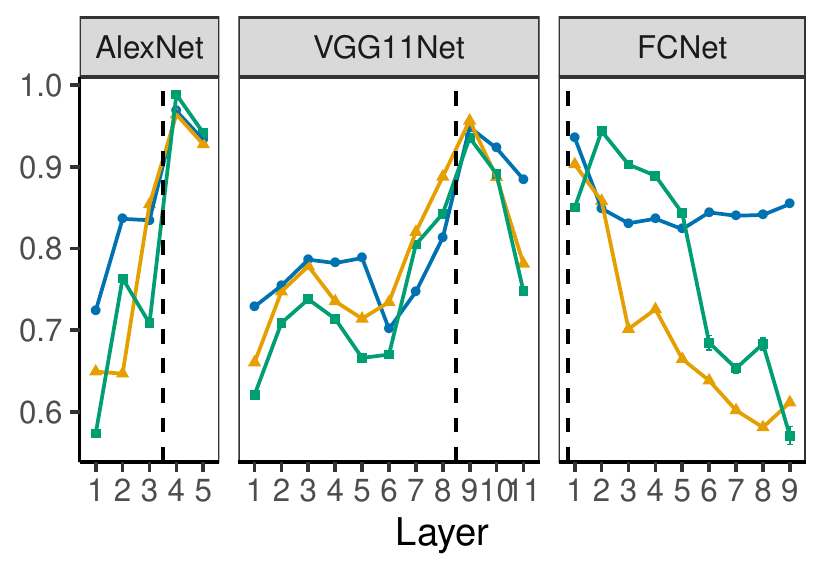}
    \caption{Results of privacy measurements \ie~(Left) $\mathcal{V}$-information, (Middle) sensitivity with $F$-norm, and (Right) AUC scores of PIA on each layer of AlexNet, VGG11Net, and FCNet, using LFW (\orangeline), CelebA (\blueline), PubFig (\greenline) dataset. Dashed lines (\blackdashedline) refer to the connection from a Conv layer to an FC layer. Each trial runs 20 times, and error bars are 95\% confidence interval.}
    \label{fig:privacy_overview}
\end{figure}

\section{Empirical characterization and validation} \label{sec:eval}
For our empirical evaluation we consider a scenario where we have a number of distributed devices, as e.g. in mobile settings or smart cities, and study the information leakage on several real-world models: variational Alexnet~\citep{krizhevsky2012imagenet} and VGG11Net~\citep{simonyan2014very}. We use Labeled Faces in the Wild (LFW)~\citep{huang2008labeled}, Large-scale CelebFaces Attributes (CelebA)~\citep{liu2015faceattributes}, and Public Figures Face Database (PubFig)~\citep{kumar2009attribute}. Due to space limitations we refer an interested reader to Appendix~~\ref{sec:evaluation_privacy} for more details.

\paragraph{Layer-wise privacy characterization.}
The $\mathcal{V}$-information and sensitivity are measured on each layer of three models, AlexNet, VGG11Net, and FCNet, trained on three datasets, LFW, CelebA, and PubFig, respectively. For the three datasets, the main task of FL is to learn `Gender', `Glasses', and `Gender', respectively, and the test accuracy using FedSGD reaches 99\%, 99\%, 94\% for LFW, 87\%, 82\%, 84\% for CelebA, and 97\%, 98\%, 92\% for PubFig. 
As shown in Figure~\ref{fig:privacy_overview}, the PIA's AUC scores have similar patterns with the prediction of both $\mathcal{V}$-information and sensitivity; \ie~the first FC layers have the highest private risk in terms of property information. More specifically, they both have a similar pattern, showing that for these three models, the $4^{th}$, $9^{th}$/$10^{th}$, and $1^{st}$ layer have the highest privacy risk respectively. %For AlexNet, the layer with the highest risk locates just after the last Conv layer (\ie~connecting to the first FC layer). This pattern is the same in the $\mathcal{V}$-information results of VGG11Net, but sensitivity indicates that the most sensitive layer is the $10^{th}$ layer which differs from the prediction of $\mathcal{V}$-information (\ie~the $9^{th}$ layer). One possible reason for this difference is that we compute the sensitivity \wrt~outputs ($\hat{\mathrm{Y}}$) instead of property ($\mathrm{p}$), so it may then move towards the last output layer. 

\begin{figure}[t] 
    \centering
    \includegraphics[width=1\textwidth]{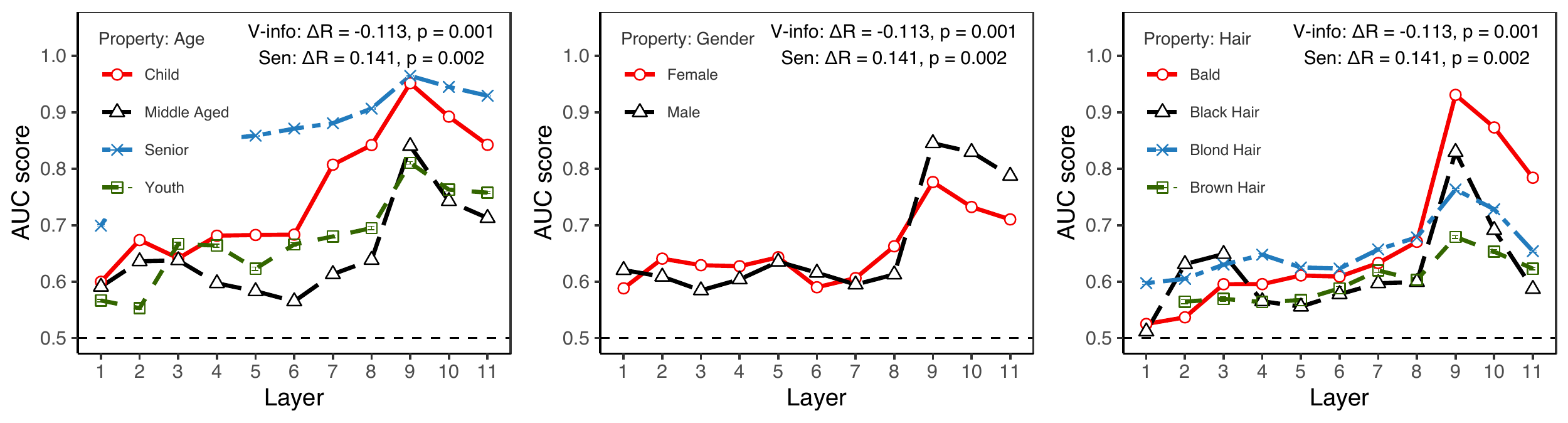}
    \caption{AUC scores of PIA aiming at (Left) Age, (Middle) Gender, and (Right) Hair on each layer of VGG11Net trained on LFW with Glasses as the main task. Pearson correlations between the AUC score and privacy metrics. Each trail runs 20 times
    %,and error bars (95\% CI) are too small to show
    .}
    \label{fig:pia_res}
\end{figure}

\begin{figure}[t]
    \centering
    \includegraphics[width=0.55\textwidth]{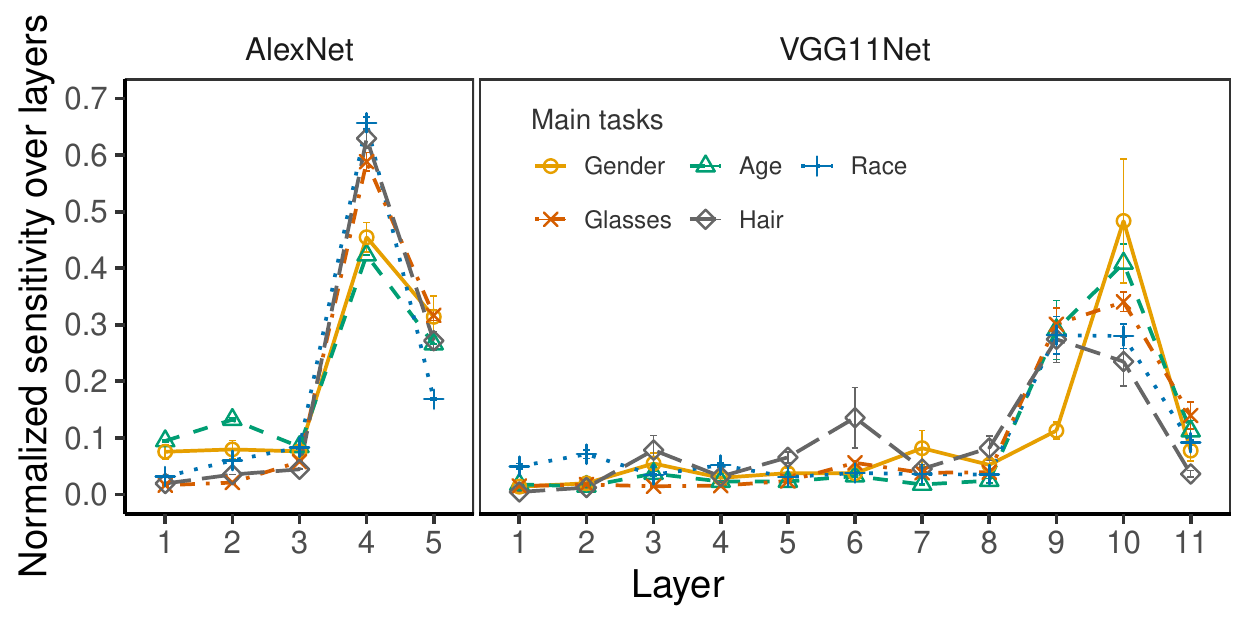}
    \vspace{-5pt}
    \caption{Normalized sensitivity ($F$-norm) over all layers of one model. Each trail runs 20 times.
  %, and error bars are 95\% CI (some are too small to show).
  }
\label{fig:sen_multitask}
\end{figure}

\paragraph{Validation on multiple properties.}
Figure~\ref{fig:pia_res} illustrates the cases where the main task of FL is the classification of `Glasses', and the PIA aims to infer properties of \emph{Age}, \emph{Gender}, and \emph{Hair}. We observe similar patterns for PIAs on all properties. That is, the $9^{th}$ layer, \ie~the first FC layer after the last Conv layer in VGG11Net, still leaks the largest amount of property information. Furthermore, the correlation coefficient ($R$ with significance level $p$) between $\mathcal{V}$-information/Sensitivity and attack AUC scores are given in the Figure, and the $\Delta R$ refers to coefficient changes when measuring other properties (\ie~AUC scores in Figure~\ref{fig:privacy_overview}). We observe a decreasing prediction capability of $\mathcal{V}$-information on the other property, and an increased capability of sensitivity. 
Figure~\ref{fig:sen_multitask}, trained on LFW, shows that no matter what the main classification task is, the first FC layers always have the highest (normalized) sensitivity. Moreover, we observe that, with the first FC layer of both networks (layer $4^{th}$ of AlexNet and layer $9^{th}$ of VGG11Net), `Gender' has a lower sensitivity than the others, while in the next FC layer, situation changes; `Gender' has a high sensitivity. As `Gender' is a more abstract feature than `Hair' color or whether `Glasses' exist, the results indicate that a more high-level property tends to be more sensitive in latter layers compared to other properties.
%and `Age' have lower sensitivity than the others in layer; with VGG11Net, they have higher sensitivity than the others in layer $10^{th}$, but `Gender' has the lowest sensitivity in layer $9^{th}$. 
%\ana{is this clear?!} \vincent{yes! thanks} %It is shown that different attributes have different distributions of sensitivity among layers and models.

% \vspace{-0.5cm}

% \vspace{-0.5cm}
\section{Conclusion and future work}
\label{sec:conclusion}
In this work, we proposed two mathematically-motivated metrics, $\mathcal{V}$-information and sensitivity, and showed that these allow to quantify latent information leakages in FL. We preliminarily validated their performance on layer-wise privacy characterization.

% maybe quantify or mention the key contribution/enabler
In future work we aim to explore i) applying the metrics on other types of datasets including time series or texts, ii) settings considering different levels of aggregations such as with more participating clients, iii) how the metrics could benefit the design of a defense mechanism leveraging layer-based privacy measurements (\eg~\cite{mo2019efficient, mcmahan2018general}). In combination with the proposed metrics, we will provide a better understanding of \emph{when} a model can leak sensitive information in FL and reveal opportunities to design flexible defenses for a better trade-off between privacy guarantees and cost.

%\newpage
\bibliography{references}
\bibliographystyle{iclr2021_conference}

\appendix
\section{Appendix}

\subsection{Notations}
\label{sec:notations}
Throughout this paper, we use lower-case italic, \eg~$x,y$, for deterministic scalar values; lower-case bold italic, \eg $\bm{x},\bm{y}$, for deterministic vectors; upper-case bold italic, \eg $\bm{X}, \bm{Y}$, for deterministic matrices. We use lower-case normal, \eg~$\mathrm{x}, \mathrm{y}$, for instances of a random variable, and upper-case normal, \eg~$\mathrm{X}, \mathrm{Y}$, for random variables of any dimensions.

\subsection{Gradient computation}
\label{sec:updated_gradients}
Here we give more background on the gradient computations and specifically how the \emph{backward} pass can reveal sensitive information. 

We first revisit how gradients are computed as a function of the input to which later shows the root cause for how a gradient $\bm{G}$ can reveal input data $\bm{X}$ and information about the property $\mathrm{p}$. Let $\bm{X} = [\bm{x}_1, \dots, \bm{x}_K]$ 
be one batch of data consisting of $K$ samples from client $c$'s training dataset $\mathcal{X}^c$, and let $\bm{Y} = [\bm{y}_1, \dots, \bm{y}_K]$ be the corresponding ground truth (\eg~labels for a classification task). The complete training dataset of all $C$ clients is denoted by $\mathcal{X} = \{\mathcal{X}^1,...,\mathcal{X}^C\}$. For a DNN model with $L$ layers, we denote layer $l$'s parameters with $\bm{W}_l$ and $\bm{b}_l$; the weights and biases, respectively. In the \emph{forward propagation} from layer $1$ to $L$, the layer $l$ computes 
$$\bm{A}_l = [\bm{a}_{l,1}, \dots, \bm{a}_{l,K}] = \bm{W}_l \bm{T}_{l-1} + \bm{b}_l \bm{1}^\mathsf{T},$$
where $\bm{W}_l \in \mathbb{R}^{N_l \times N_{l-1}}$, $\bm{T}_{l-1} = [\bm{t}_{l-1,1}, \dots, \bm{t}_{l-1,K}] \in \mathbb{R}^{N_{l-1}\times K}$, $\bm{b}_l \in \mathbb{R}^{N_l}$, $\bm{1} \in \{1\}^{K}$. $N_l$ denotes the size (\ie~the number of neurons) of the layer $l$, and $\bm{T}_{l-1}$ shows the \emph{intermediate representation} that is the output of the (previous) layer $l-1$, and $\bm{t}_{l-1,k}$ corresponds to one sample $k$'s outputs. Then, the layer $l$'s output is $\bm{T}_l = [\sigma (\bm{a}_{l,1}), \dots, \sigma(\bm{a}_{l,K})] \in \mathbb{R}^{N_{l}\times K}$, denoted as $\sigma (\bm{A}_l)$ for simplicity, where $\sigma(\cdot)$ is the chosen activation function. Note that $\bm{T}_{0} = \bm{X}$ and $\bm{T}_{L} = \bm{\hat{Y}}$ the prediction on ground truth.

In the \emph{backward propagation}, the loss $\ell$ between $\hat{\bm{Y}}$ and $\bm{Y}$ propagates from the layer $L$ to $1$. For layer $l$, the gradient vector $\bm{G}_l$ consists of the gradients of the weights and biases $\{ \bm{G}^{(w)}_l, \bm{G}^{(b)}_l\}$, which are computed using chain rule:
\begin{equation}
    \bm{G}^{(w)}_l = 
    \frac{\partial \ell}{\partial \bm{W}_l} =
    \frac{\partial \ell}{\partial \bm{A}_{l}}
    \frac{\partial \bm{A}_{l}}{\partial \bm{W}_l}
    =
    \frac{\partial \ell}{\partial \bm{A}_{l}} \bm{T}^{\mathsf{T}}_{l-1}
    \label{eq:gradient_weights}
\end{equation}
\begin{equation}
    \bm{G}^{(b)}_l = 
    \frac{\partial \ell}{\partial \bm{b}_l} =
    \frac{\partial \ell}{\partial \bm{A}_{l}}
    \frac{\partial \bm{A}_{l}}{\partial \bm{b}_l}
    =
    \frac{\partial \ell}{\partial \bm{A}_{l}} \bm{1}
    \label{eq:gradient_biases}
\end{equation}
where $\bm{G}^{(w)}_l \in \mathbb{R}^{N_l \times N_{l-1}}$ and $\bm{G}^{(b)}_l \in \mathbb{R}^{N_l}$ and have the same size with $\bm{W}_l$ and biases $\bm{b}_l$ respectively.

The set of all layers' $\bm{G}^{(w)}$ and $\bm{G}^{(b)}$ in Equation~\ref{eq:gradient_weights} and~\ref{eq:gradient_biases} is the minimum unit to update in FL, \ie~corresponding to the FedSGD. $\bm{G}^{(w)}$ and $\bm{G}^{(b)}$ are also targeted by adversaries to extract private information of $\bm{X}$ or its sub-information \eg~$\mathrm{p}$. In FedAvg, before updating, more batches of data are fed for training, and the gradients of multiple batches are aggregated together element-wisely by
\begin{equation}
    \sum_{\bm{X}_i \in \mathcal{X}^c} \{\bm{G}^{(w)}_l\}_{\bm{X}_i} ; \; \sum_{\bm{X}_i \in \mathcal{X}^c} \{\bm{G}^{(b)}_l\}_{\bm{X}_i}
\label{eq:gradient_sum}
\end{equation}

\subsection{General analysis on gradients' privacy risks}
\label{sec:privacy_x}

We first directly analyze how $\bm{G}$ leaks information of $\bm{X}$ or $\mathrm{p}$ based on Equation~\ref{eq:gradient_weights},~\ref{eq:gradient_biases}, and~\ref{eq:gradient_sum}. Three \emph{indications} for information leakage from DNN parameters (\ie~gradients) are: (1) One can obtain the layer $l$'s input ($\bm{T}_{l-1}$) based on this layer's gradient updates. Because the first layer's input (\ie~$\bm{T}_0$) is $\bm{X}$, the original input is highly likely to be leaked, which has been shown in previous research~\citep{aono2017privacy}.
(2) Batch-based SGD (\ie~Minibatch) reduces the potential leakage on one specific data sample $\bm{x}_i, i \in 1,...,K$, because gradients are computed over all samples in one batch (see Equation~\ref{eq:gradient_weights} and~\ref{eq:gradient_biases}). Aggregation, \ie~the summation of gradients over multiple batches, also reduces the potential leakage (see Equation~\ref{eq:gradient_sum}).
(3) The term $\frac{\partial \ell}{\partial \bm{A}_l} = \frac{\partial \ell}{\partial \bm{T}_l} \odot \sigma' (\bm{a}_l) = \bm{W}_{l+1}^{\mathsf{T}} \frac{\partial \ell}{\partial \bm{A}_{l+1}} \odot \sigma' (\bm{A}_l)$ (note that $\odot$ is the Hadamard product), indicating that the gradient can potentially contain information of $\bm{A}_l$, $\bm{W}_{l+1}$, $\bm{b}_{l+1}$, and parameters of following layers (\ie~$\bm{A}_{l+1},...,\bm{A}_{L}$). However, the DPI (from layer 1 to $L$) indicates that $\mathrm{T}_{l+1}$ (or $\bm{a}_{l+1}$ here) cannot contain more information on $\mathrm{X}$ than $\mathrm{T}_{l}$ (or $\bm{a}_{l}$) in \emph{forward propagation}~\citep{saxe2019information, shwartz2017opening, tishby2000information}.

\begin{figure}[h!]
    \centering
    \includegraphics[width=0.5\textwidth]{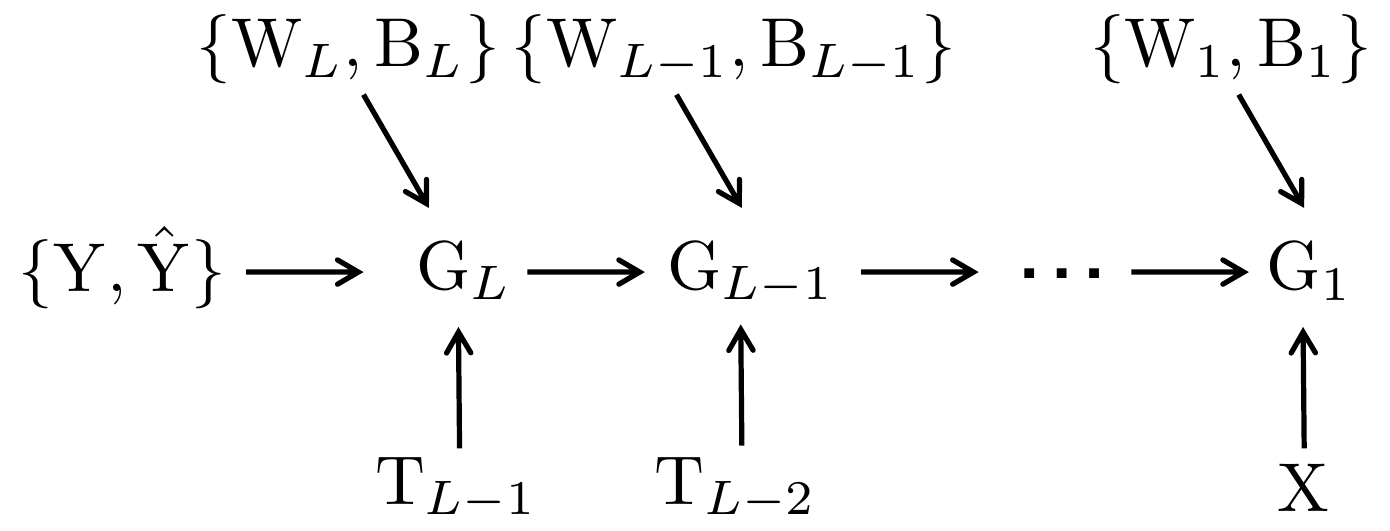}
    \caption{Markov chain from layer $L$ to layer $1$ when producing gradients ($\mathrm{G}$) based on weights ($\mathrm{W}$), biases ($\mathrm{B}$), and intermediate representations ($\mathrm{T}$) in backward propagation (Note: $\hat{\mathrm{Y}}$ is $\mathrm{T}_L$, and $\mathrm{X}$ is $\mathrm{T}_0$).}
    \label{fig:mkc}
\end{figure}

\vspace{5pt}
\noindent \textbf{Backward Markov chain}.
Similar to the Markov chain in the forward propagation, here we present the Markov chain in \emph{backward propagation} from layer $L$ to layer $1$ based on how gradients are computed to clarify the information leakage from gradients. As shown in Figure~\ref{fig:mkc}, $\mathrm{G}_l$ is produced based on layer $l$'s weights, biases, the previous representation, and next gradient (denoted by $\mathrm{W}_l, \mathrm{B}_l, \mathrm{T}_{l-1}$, and $\mathrm{G}_{l+1}$). 
Previous research~\citep{shwartz2017opening, saxe2019information} has indicated that, due to the data processing inequality, $\mathrm{T}_l$ contains more information about $\mathrm{X}$ then $\mathrm{T}_k$ with $k>l$. Furthermore, matrix $\bm{W}_l$ and $\bm{b}_l$ are linearly aggregated gradient updates from multiple previous $\bm{G}_l$ (\ie~the model parameters $\bm{\theta}_t \leftarrow \bm{\theta}_{t-1} - \bm{G}_t$ for the $t^{th}$ batch data), so intuitively $\mathrm{W}_{l}$ and $\mathrm{B}_{l}$ possibly not contain more information on $\mathrm{X}$ than $\mathrm{G}_{l}$, and then the information of $\mathrm{X}$ in $\mathrm{T}_{l-1}$ may reflect that of $\mathrm{X}$ in $\mathrm{G}_l$. 
%See Appendix~\ref{sec:info_x_t} for more analyses of estimating information on $\mathrm{T}$.

Moreover, related to indication (2), when it comes to a particular $\mathrm{x}$, $\mathrm{W}$ and $\mathrm{B}$ can be regarded as less sensitive if they have been updated based on the complete dataset $\mathcal{X}$. We further argue that the attempt on evaluating gradient leakage on original data $\mathrm{x}_i, i \in 1,...,K$, such as~\citep{wei2020framework}, could be very limited and constrained under weak aggregation, \ie~a small batch size~\citep{zhu2019deep}.% We refer to  Appendix~\ref{sec:attack_x} for data reconstruction attack results on $\mathrm{G}$, which shows the first layer could be the most sensitive in terms of original data.

\subsection{Empirical Setup}
\label{sec:evaluation_privacy}

Here, we explain our experimental evaluation set-up that has been used for the empirical layer-wise characterization and validation of the predictive ability of the proposed metrics.

\vspace{5pt} 
\noindent \textbf{Goals and validity.} 
Our first evaluation goal is to apply the proposed metrics, $\mathcal{V}$-information and sensitivity, to characterize latent information privacy on DNNs. %The results of this characterization are presented in Section~\ref{subsec:res_char}. 
Our second goal is to establish the validity of the privacy measurements.
%While our analysis is grounded in mathematical foundations we need an external or established approach to validate its results. 
We leverage the fact that the adversary's success can be an indicator of privacy risks and commonly used in previous research~\citep{yeom2018privacy,jayaraman2019evaluating,wei2020framework,mo2020darknetz}; intuitively, if an adversary has access to information of higher risk this should yield better attack results. Therefore, 
%to validate our proposed privacy metrics, $\mathcal{V}$-information and sensitivity, 
for each of the layers of a model, we compare both $\mathcal{V}$-information and sensitivity  with the \textit{PIA adversary's AUC score} (\ie~area under the ROC curve) when the adversaries have only access to that layer.  AUC score is the most common evaluation method for PIAs' success~\citep{powers2011roc, melis2019exploiting}. %The results of PIAs validating the metrics are presented and discussed in Section~\ref{subsec:res_char_one} and~\ref{subsec:res_char_multi}. 
In addition, parameter settings of metrics such as the \emph{chosen attack model} in $\mathcal{V}$-information and \emph{norms} in sensitivity analysis may influence the evaluation results, so they are also measured together in our validations.

\vspace{5pt}
\noindent \textbf{Models.}
For the characterization and validation of the layer-wise privacy risk, we first use two DNN architectures used in~\citep{melis2019exploiting}: variational AlexNet and VGG11Net. Both consist of several convolutional (Conv) layers as the earlier layer followed by several fully connected (FC) layers as the latter layers. This type of architecture is widely used because of its good performance; Conv layers capture features of the input data from low to high levels and then following FC layers process classification on them. Specifically, our AlexNet has three Conv layers with 16, 32, 64 filters of $3\times3$ size (and each has a max-pooling with a size of $(2,2)$ after), followed by two FC layers with 256 and $d_y$ neurons (where $d_y$ is the output size or the number of classes). VGG11Net has eight Conv layers with 16, 16, 16, 16, 32, 32, 64, 64 filters of $3\times3$ size (each of the last three has a max-pooling with a size of $(2,2)$ after), followed by three FC layers with 256, 128, $d_y$ neurons, respectively. In addition, we include another fully connected network (FCNet) which has nine FC layers. All these FC layers have 32 neurons except the output (last) layer. The FCNet architecture has been used in previous research~\citep{shwartz2017opening, saxe2019information}; by evaluating FCNet we aim to investigate the privacy risk when only FC layers are presented and when layers have the same size but different locations among the model. All three DNNs use ReLU activation functions for all layers (except the output layer).

\vspace{5pt} \noindent \textbf{Datasets.} The three models described above are trained on three datasets with attributes, including Labeled Faces in the Wild (LFW)~\citep{huang2008labeled}, Large-scale CelebFaces Attributes (CelebA)~\citep{liu2015faceattributes}, and Public Figures Face Database (PubFig)~\citep{kumar2009attribute}. LFW contains 13233 face images cropped and resized to $62\times47$ RGB. All images are labeled with attributes such as gender, race, age, hair color, and more. CelebA contains more than 200k face images of celebrities with 40 attribute annotations such as gender, hair color, eyeglasses, and more. We use a subset of the cropped version (\ie~15000 images) and resize images to $64\times64$ RGB. We also use a cropped version ($100\times100$ RGB) of PubFig which contains 8300 facial images made up of 100 images for each of 83 persons~\citep{pinto2011scaling}. All images are marked with 73 attributes, \eg~gender and race.

\vspace{5pt}
\noindent \textbf{Simulation.} We conduct our experiments on a cluster with multiple nodes where each has 4 Intel(R) Xeon(R) E5-2620 CPUs (2.00GHz), an NVIDIA RTX 6000 GPU (24GB), and 24GB DDR4 RAM. Deep learning framework Pytorch~\citep{paszke2019pytorch} v1.4.0 is used for privacy measurements, and Theano~\citep{bergstra2010theano} v1.0 is used for PIAs. We follow the FL setting in the seminal PIA paper~\citep{melis2019exploiting} for comparing the privacy measurements with PIA results. Specifically, we conduct the learning process using FedSGD as the optimization algorithm. The learning rate is set as 0.01 without momentum. The batch size is 32. The training datasets above are partitioned into two parts to simulate two clients, where one client is assumed as the adversary. As mentioned in~\citep{melis2019exploiting}, similar results can be achieved for more clients.  The adversary first participates in the training with others for several communication rounds (\eg~100 rounds) and saves snapshots of the received global model to conduct PIAs later. For measuring the ultimate privacy cost, we compute $\mathcal{V}$-information and sensitivity at the end of total communication rounds.

\end{document}